\documentclass[namedreferences]{solarphysics}
\usepackage[optionalrh]{spr-sola-addons} % For Solar Physics 
\usepackage{graphicx}        % For eps figures, newer & more powerfull
\usepackage{natbib}         % For citations: redefine \cite commands
\usepackage{color}           % For color text: \color command
\usepackage{url}             % For breaking URLs easily trough lines
\usepackage[outdir=./]{epstopdf}
\usepackage{enumerate}
\newcommand{\paperi}{Paper I}

\newcommand{\DN}{DN s$^{-1}$ pix$^{-1}$}

\begin{document}

\begin{article}

\begin{opening}

\title{GRID-SITES: Gridded Solar Iterative Temperature Emission Solver for Fast DEM Inversion}

\author{James~\surname{Pickering}}
\author{Huw~\surname{Morgan}\**}
\runningauthor{J. Pickering and H. Morgan}
\runningtitle{Grid-SITES}

   \institute{$^{1}$ Physics Department, Aberystwyth University, Ceredigion, Cymru, SY23 3BZ\\
                     email: \url{jap39@aber.ac.uk, hmorgan@aber.ac.uk} \\ 
             }

\begin{abstract}
The increasing size of solar datasets demands highly efficient and robust analysis methods. This paper presents an approach that can increase the computational efficiency of differential emission measure (DEM) inversions by an order of magnitude or higher, with the efficiency factor increasing with the size of the input dataset. The method, named the Gridded Solar Iterative Temperature Emission Solver (Grid-SITES) is based on grouping pixels according to the similarity of their intensities in multiple channels, and solving for one DEM per group. This is shown to be a valid approach, given a sufficiently high number of grid bins for each channel. The increase in uncertainty arising from the quantisation of the input data is small compared to the general measurement and calibration uncertainties. In this paper, we use the Solar Iterative Temperature Emission Solver (SITES) as the core method for the DEM inversion, although Grid-SITES provides a general framework which may be used with any DEM inversion method, or indeed any large multi-dimensional data inversion problem. The method is particularly efficient for processing larger images, offering a factor of 30 increase in speed for a 10 megapixel image. For a time series of observations, the gridded results can be passed sequentially to each new image, with new populated bins added as required. This process leads to increasing efficiency with each new image, with potential for a $\approx$100 increase in efficiency dependent on the size of the images. 
\end{abstract}
\end{opening}
%-------------------------------------------------

%The \textit{Atmospheric Imaging Assembly} on the \textit{Solar Dynamics Observatory} (AIA/SDO) collects data at a rate that makes a complete temperature and emission (or a differential emission measure, DEM), analysis close to impossible without high-performance computing. 

\section{Introduction}
\label{intro}

Extreme ultraviolet (EUV) observations are a crucial source of information on the low solar corona. Qualitatively, EUV images provide a window into the structure and dynamics of the atmosphere. Quantitatively, both EUV images and spectroscopy constitute the most important information of the plasma characteristics of the transition region and low corona. A major part of this information is provided by differential emission measure (DEM) analysis. A DEM is a powerful diagnostic of the coronal plasma - it is an estimate of the emission, or the total number of electrons squared along the observed line of sight (similar to a column mass), at a given temperature. EUV measurements in multiple bandpasses (or spectral emission lines) with different temperature responses, allow the estimation of a DEM. DEM analysis is core to many important studies of various solar atmospheric features and events, including coronal mass ejections in their initial stage of eruption \citep{cheng2012,hannah2013}, active regions and loops \citep{aschwanden2013, warren2012, delzanna2013}, flares \citep{fletcher2013,dudik2014,sun2014,su2018,hernandez2019}, EUV waves \citep{kozarev2011,vanninathan2015}, and coronal dimmings \citep{vanninathan2018,veronig2019}.

As new EUV instruments are developed, the temporal, spatial and spectral resolution becomes ever finer. Since 2010, the \textit{Atmospheric Imaging Assembly} (AIA: \opencite{lemen2012}) aboard the \textit{Solar Dynamics Observatory} (SDO: \opencite{pesnell2012}) provides very fine temporal and spatial resolution of the Sun at multiple wavelengths. Even as the community continues to develop methods to digest the large volume of data from AIA/SDO, new instruments are planned and tested with even finer resolution (\emph{\emph{e.g.}} \textit{High-Resolution Coronal Imager} (Hi-C), see \opencite{cirtain2013}). In the context of the size of the AIA/SDO dataset, or the data collection rate, DEM inversion methods are computationally expensive. Larger DEM studies of broad regions over long time periods are rare (\emph{\emph{e.g.}} \opencite{morgan2017}). 

To our knowledge, the most computationally fast method is that of \citet{cheung2015}, based on Simplex optimization of a set of smooth basis functions, or a sparse matrix. This method can process around $10^5$ pixels per second on a standard desktop computer. The Solar Iterative Temperature Emission Solver (SITES) inversion method (\opencite{morgan2019}, hereafter \paperi) creates an  initial  DEM  estimate  through  a  direct  redistribution  of  observed  intensities across temperatures according to the temperature response function of the measurement, and iteratively improves on this estimate through calculation of intensity residuals. The resulting DEMs replicate model DEMs well in tests on AIA synthetic data. It is simple in concept and implementation and is non-subjective in the sense that no prior  constraints are placed on the solutions other than positivity and smoothness, and can process a 1000 DEMs per second on a standard desktop computer. This is similar to the speed of regularized matrix inversion-based methods such as \citet{hannah2012} or \citet{plowman2013}. 

This paper provides a general framework for improving the efficiency of applying DEM inversions to data, with emphasis on AIA images. It is not a new DEM inversion method, but an approach to pre-sorting input measurements to enable faster DEM inversion. Thus Grid-SITES may be used with any inversion method, although SITES is used throughout this work as the core inversion method. See \paperi\ for a description of SITES and an overview of other DEM inversion methods.

The Grid-SITES concept and method is presented in Section \ref{gridconcept}, and its validity confirmed in Section \ref{valid}. Section \ref{uncertain} describes the additional uncertainty introduced by the gridding method. Results, and comparison of Grid-SITES to directly-inverted DEMs are shown in Section \ref{results}. Section \ref{timeseries} describes the application of Grid-SITES to a time series of observations, and Section \ref{fulldisk} shows the inversion of full-resolution AIA observations. A brief summary is given in Section \ref{summary}.

\section{The Gridding Concept and Method} %%%%%%%%%%%%%%%%%%%%%%%%%%%%%%%%%%%%%%%%
\label{gridconcept}

The gridding concept assumes that many of the pixels in an AIA observation will have a very similar set of intensities across all channels, and further assumes that these similar intensities will result in similar DEMs. Both these assumptions are shown to be valid in this section, offering a great saving in computational efficiency through calculating only one DEM for each group of pixels which are deemed similar. Given the relatively high calibration uncertainties \citep{boerner2014}, in addition to the errors arising from the DEM inversion, the added uncertainty of grouping non-identical, yet similar measurements together to give a single DEM is shown to be reasonable, and adds only a small additional error to the resulting DEM.

\begin{figure}    %%%%%%%%%%%%%%%%%% FIGURE 1 
   \centerline{\includegraphics[width=0.95\textwidth,clip=]{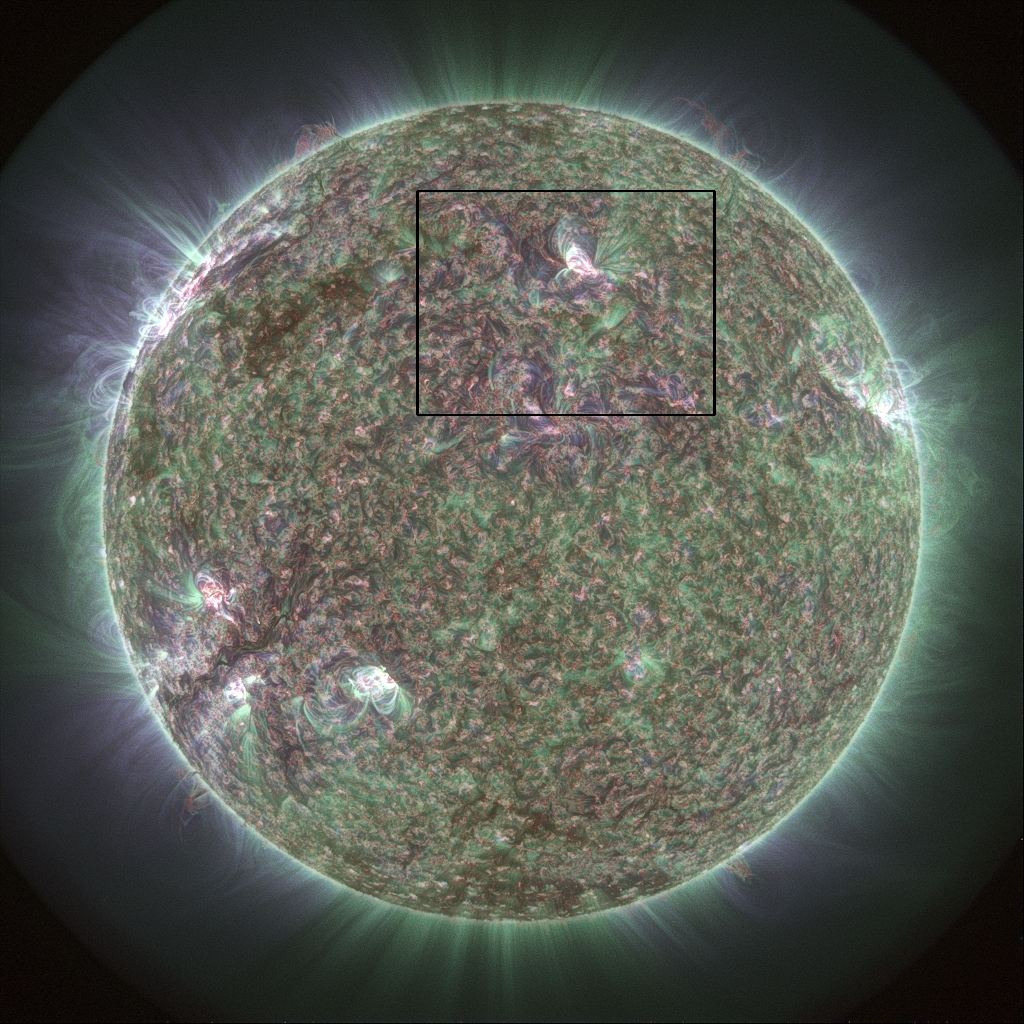}}
   \caption{A context image from 01 January 00:00UT, created from a set of AIA `synoptic' observations. The boxed region is used to demonstrate the method. All seven AIA channels contribute to this composite, with the temperature response of each channel between 0.05 and 7.0MK specifying that channel's contribution to the red, green and blue colour channels of the output images. The image is processed with multiscale Gaussian normalization to enhance fine-scale structure \citep{morgan2014}. Note that this is a processed image provided for spatial context, thus is not directly representative of the observed numerical values used for the analysis.}
    \label{aiaimage}
  \end{figure}

A set of AIA observations made at 01 January 00:00UT are used to introduce the gridding concept and method. Six EUV channels are included: 94, 131, 171, 193, 211 and 335\,\AA. For convenience, the `synoptic' images are used (\url{http://jsoc2.stanford.edu/data/aia/synoptic/}), that are reduced in spatial dimensions to $1024\times1024$ pixels through averaging, thus 16 pixels in the original high-resolution data are combined to create one synoptic pixel. Figure \ref{aiaimage} shows a colour composite image of this observation set to give context. The boxed region enclosing pixels $x=[350,700]$ and $y=[600,850]$ is used as a case study. The region contains a small active region and quiet Sun. In order to test the validity of gridding, a direct DEM inversion is applied to each pixel, using the SITES method of \paperi. 

\begin{figure}    %%%%%%%%%%%%%%%%%% FIGURE 1 
   \centerline{\includegraphics[width=0.95\textwidth,clip=]{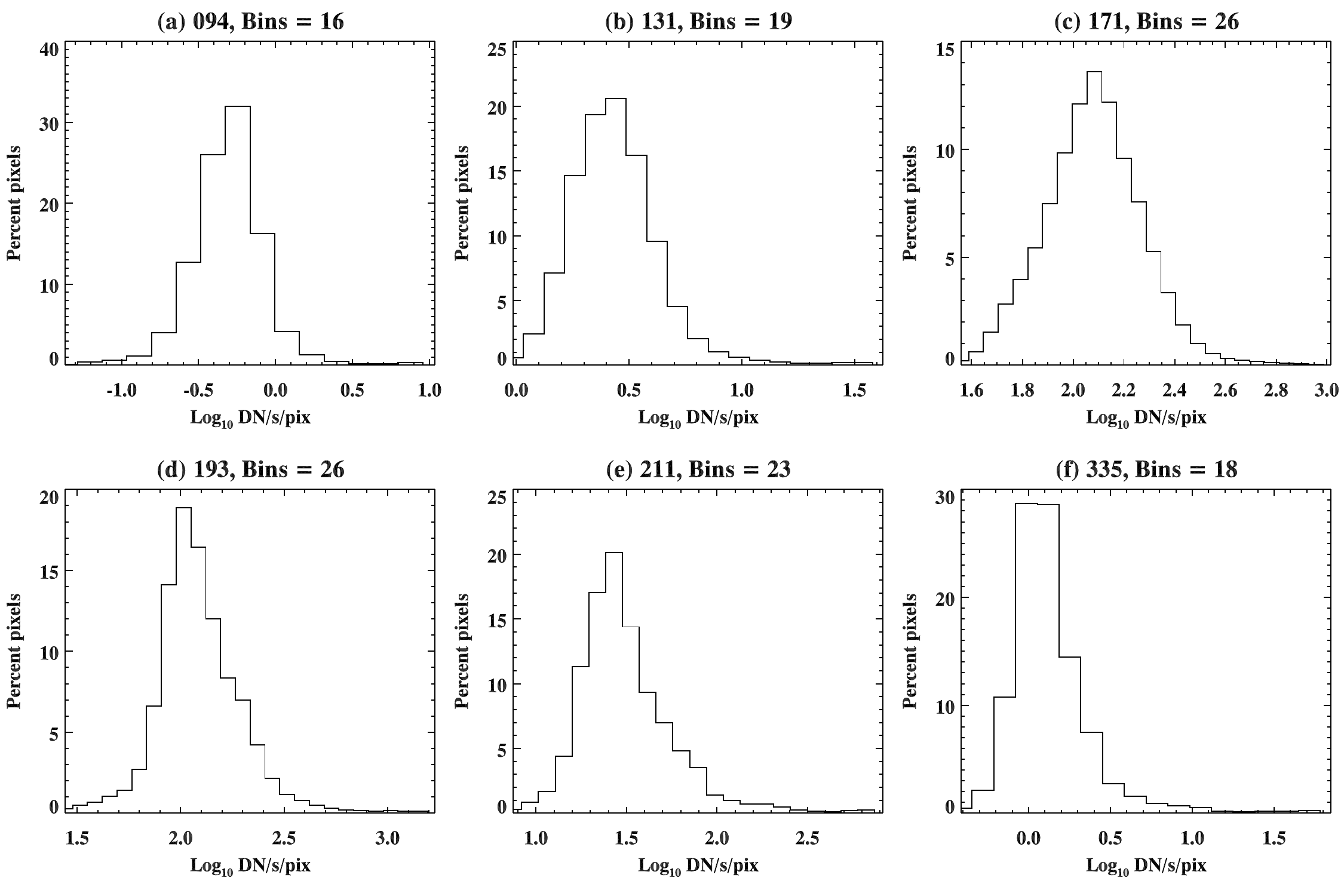}}
   \caption{Histograms of the logarithmic intensity of each AIA channel for values between the 0.1 percentile minimum and maximum values as described in the text. The parameters for each histogram are listed in Table \ref{histogram1table}}
    \label{histogram1}
  \end{figure}

Of the $351\times251=88101$ pixels in the region, a few are discarded due to spurious high ($>1.4 \times 10^4$ DN) or low ($<-5$ DN) values in one or more channel. Of the remaining pixels, the intensity in each channel is converted to a logarithmic (base ten) intensity, and the minimum and maximum 0.1\% percentile calculated. These are listed in Table \ref{histogram1table}, along with the percentage of pixels falling outside the minimum-maximum range. Figure \ref{histogram1} shows the distribution of log intensities for each channel. The number of grid bins for each channel, $n_i$, (with each channel indexed by subscript $i$) is set between a minimum of $n_0=16$ and a maximum of $n_1=26$ according to each channel's median log intensity, $\tilde{I_i}$, through
\begin{equation}
n_i = \lceil n_0 + \frac{\tilde{I_i} - \tilde{I}_{\rm max}}{\tilde{I}_{\rm max}-\tilde{I}_{\rm min}}(n_1-n_0) \rceil, 
\end{equation}
where $\tilde{I}_{\rm min}$ and $\tilde{I}_{\rm max}$ are the minimum and maximum median intensity over all channels respectively, and the curtailed square brackets represent rounding up to the nearest integer. Thus the lowest intensity channel has 16 bins and the highest has 26.

\begin{center}
\begin{table}
\begin{tabular}{ c | c | c | c | c | c }
{Channel} & {Min} & {Max} & {Number bins} & {Binsize} & {\% outliers}\\
\hline
{094} & {-1.38} & {1.04} & {16} & {0.15} & {0.36}\\
{131} & {-0.02} & {1.63} & {19} & {0.09} & {0.37}\\
{171} & {1.56} & {3.02} & {26} & {0.06} & {0.36}\\
{193} & {1.44} & {3.23} & {26} & {0.07} & {0.36}\\
{211} & {0.87} & {2.92} & {23} & {0.09} & {0.36}\\
{335} & {-0.44} & {1.85} & {18} & {0.13} & {0.37}\\
\end{tabular} 
\caption{Parameters for the histogram binning of the AIA channels (used to create Figure \ref{histogram1}), showing the 0.1 percentile minimum and maximum log intensity (\DN), number of bins, the width of each bin in logarithmic intensity, and the percentage of pixels outside the histogram range.}
\label{histogram1table}
\end{table}
\end{center}

Creating a six-dimensional histogram of the six channel intensities using standard histogram programs would result in an array of several tens to hundreds of millions of elements (depending on the choice of the number of bins in each channel). The number of bins presented in Table \ref{histogram1table} would give a histogram with over $85\times 10^6$ bins (the product over all channels of the number of bins). However, for any observation most of these elements have no members. For the example dataset used here, only $\approx2\times 10^4$ elements are populated. That is, if a six-dimensional histogram was constructed using the parameters of Table \ref{histogram1table}, it would have over $85\times 10^6$ elements, but only $\approx2\times 10^4$ elements would be populated based on the input intensities. For any AIA observation, only a relatively small number of unique intensity groupings, or populated bins, exist.  The number of populated bins for the example observation ($\approx2\times 10^4$) is around 20\%\  of the total number of pixels in the observation ($\approx8.8\times 10^4$), which suggests that a $\approx5$-fold increase in efficiency is possible if DEM inversions are applied only to the populated bins. Whilst this is not a huge efficiency increase, larger images offer greater efficiency. For example, including all the pixels on the disk for this `synoptic' observation ($\approx$0.5 megapixels) results in $\approx5\times10^4$ groups (or populated bins), which offers a $\approx$10-fold efficiency increase. For a full resolution image of the same observation date, there are $\approx$8.5 megapixels on the disk, giving $\approx10^5$ groups - a 100-fold increase in speed compared with processing each pixel separately. This increase in efficiency with a greater number of input pixels can be explained by considering the distributions of intensities in each channel. Plotting histograms of intensities for, say, the 171\,\AA\ channel for the small example region gives the distribution of Figure \ref{histogram1}c. The histogram for a full-disk image for the same channel would show a similar distribution, albeit with a higher number of pixels per bin. The same argument will hold for a six-dimensional histogram over all channels. Since the gridding scheme will only seek to calculate one DEM per populated bin, increased efficiency is achieved.

An efficient way to find these unique groupings for a large number of pixels, without creating a large histogram, is: 
\begin{enumerate}[i)]
\item Channels are indexed with subscript $i$. Each channel's intensities are converted to logarithmic intensity, $I_i$. This gives a more even distribution of values compared to using intensities directly. 
\item Establish each channel's minimum, $m_{0i}$, and maximum ,$m_{1i}$, logarithmic intensity as well as the number of bins $n_i$ for each channel. We use a robust minimum and maximum, and a higher number of bins is allocated to higher-signal channels as described above in the context of Figure \ref{histogram1} and Table \ref{histogram1table}.
\item Interpolate the intensity value of each pixel into the appropriate bin index for that channel using
\begin{equation}
k_i = \left \lfloor{ n_i \frac{(I_i-m_{0i})}{(m_{1i}-m_{0i})}}\right \rfloor.
\end{equation}
The truncated square brackets indicate rounding down to the nearest integer, thus $k_i$ gives the bin number for each measurement for that channel. The intensity at a given pixel and channel is transformed into the appropriate bin index $k_i$ for that pixel and channel, according to the number of bins, $n_i$, and minimum/maximum intensity values.
\item Initialise a main index, $K$, equal to $k_5$ (the bin number index of the last (sixth) channel in use), and repeat for each channel from $i_4$ (the fifth channel) down to $i_0$ (the first channel):
\begin{equation}
K^{\rm now}= K^{\rm prev} n_i + k_i, 
\end{equation}
thus converting the $k_i$ into a one-dimensional index giving the location of each pixel within a multi-dimensional histogram. The $k_i$ give the one-dimensional bin number at each channel (thus range from 0 to $n_i-1 = $\emph{\emph{e.g.}} $16$), whilst $K$ gives the overall bin number in the multi-dimensional histogram (thus can range from 0 to \emph{e.g.} $8.5\times 10^6$ for the small example region). Note that the choice of allocation of each histogram dimension to a particular AIA channel is not important, as long as this order remains consistent throughout the processing. We simply use increasing wavelength to allocate channels to each dimension.
\item Finally, the unique values of $K$ are identified. These are extracted to give the subset $K_u$, or the set of unique combinations of channel intensities.
\end{enumerate}

\begin{figure}    %%%%%%%%%%%%%%%%%% FIGURE 1 
   \centerline{\includegraphics[width=0.95\textwidth,clip=]{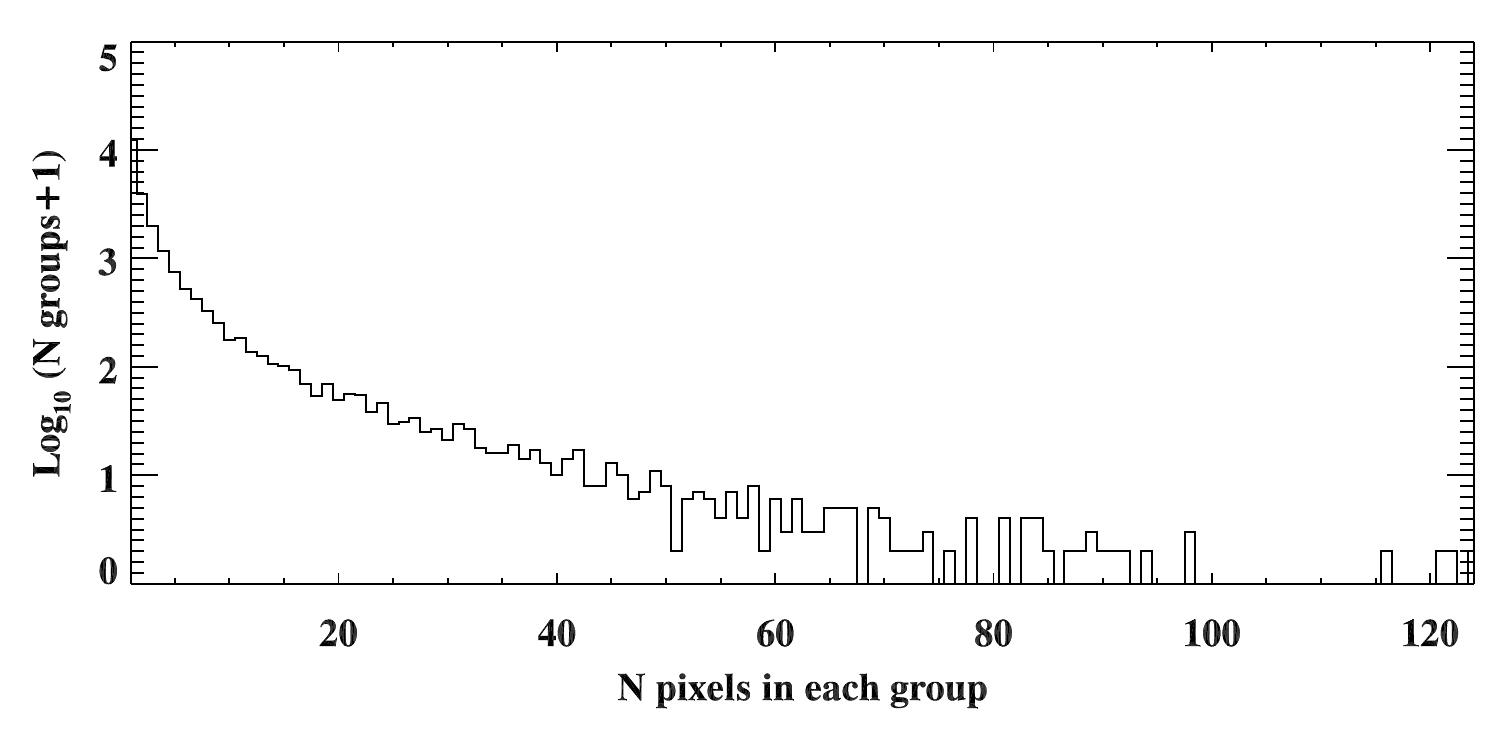}}
   \caption{Distribution of the number of groups that contain a certain number of pixels. For example, there are $>10^4$ groupings containing only one pixel, and only one group containing the maximum 124 pixels. The $y$-axis has a logarithmic scale, so a constant of one has been added to the number of groups to avoid numerical error.}
    \label{histogram2}
  \end{figure}

For the example region, there are 23,581 elements to $K_u$. Figure \ref{histogram2} shows the number of groups against the number of pixels in each group. Around half the members of $K_u$ contain only one contributing pixel. The majority of elements of $K_u$ contain only a small number of pixels, below 10, with a few groups containing over 60 pixels. The maximum number of pixels in a group is 124. 

\section{Validity of gridding}
\label{valid}

\begin{figure}    %%%%%%%%%%%%%%%%%% FIGURE 1 
   \centerline{\includegraphics[width=0.95\textwidth,clip=]{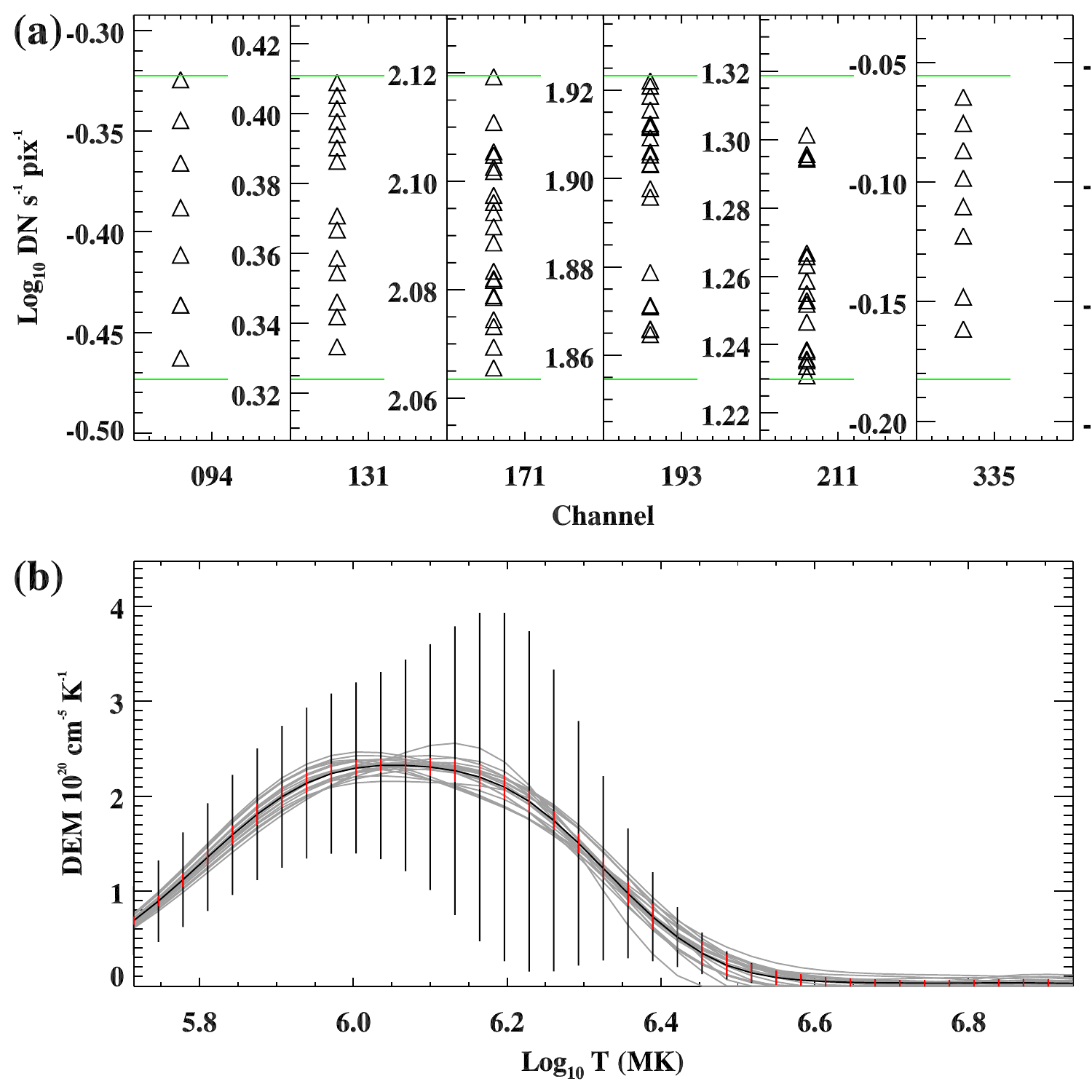}}
   \caption{(a) The intensities in each channel for a group of 20 similar pixels are shown here as black triangles. The green lines show the range of each intensity histogram bin which encompass the measured intensities. (b) The set of DEMs calculated directly for each pixel within this group are shown as grey lines. The mean of these is shown as a bold black line, with their standard deviation as red error bars. The black error bars gives the DEM error estimated from error propagation of the input measurements errors and the AIA calibration uncertainties (see text).}
    \label{demvar}
  \end{figure}

We wish to test the assumption that grouping similar pixels, according to the set number of intensity bins, give rise to similar DEMs. An arbitrary group containing 20 similar pixels is chosen, and each pixel's channel intensities are used to calculate DEMs directly using the method of \paperi. Figure \ref{demvar}a shows the small range of each channel's intensities, and their correct correspondence to the intensity binning ranges. Figure \ref{demvar}b shows the 20 individual DEMs as grey lines. They are very closely distributed, reflected in their small standard deviation shown as red error bars. The mean DEM is shown as a bold black line, and the black error bars show the estimated uncertainty in the DEM, calculated from the measurement errors and calibration uncertainties according to Equation 6 of \paperi. The spread of the 20 DEMs is far smaller than the estimated DEM uncertainty, showing that for this group, the assumption is valid. 

%Figure \ref{demvar
%
%
%\begin{figure}    %%%%%%%%%%%%%%%%%% FIGURE 1 
%\centerline{\includegraphics[width=0.95\textwidth,clip=]{demvarmax.pdf}}
%\caption{As figure \ref{demvarmax}, but for the group containing 124 pixels (largest group).}
%\label{demvarmax}
%\end{figure}

\begin{figure}    %%%%%%%%%%%%%%%%%% FIGURE 1 
   \centerline{\includegraphics[width=0.95\textwidth,clip=]{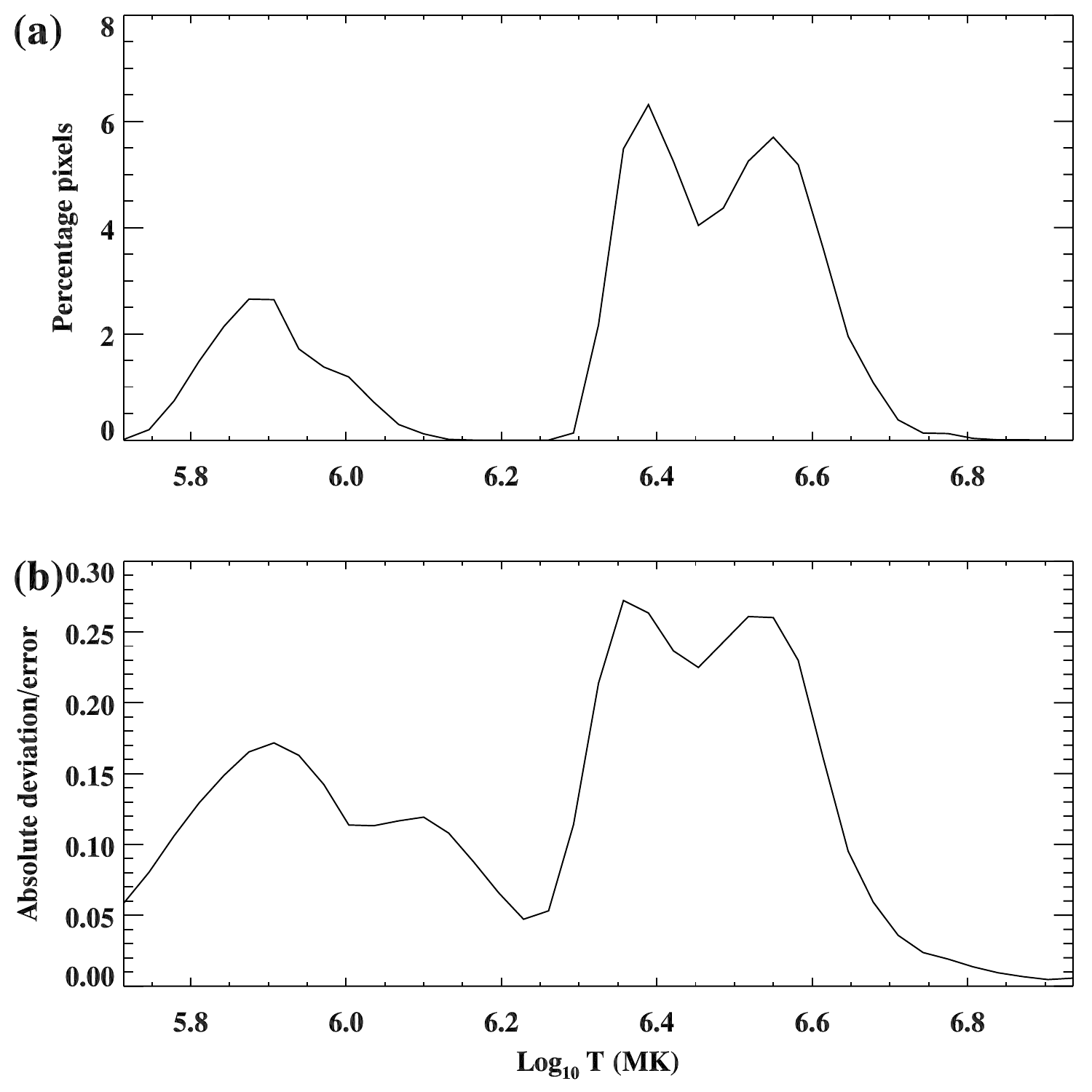}}
   \caption{(a) The percentage of pixels, at each DEM temperature bin, that have Grid-SITES output DEMs that lie outside the range of the estimated DEM inversion uncertainty as calculated by direct DEM inversion of each pixel separately. This deviation beyond the inherent inversion uncertainty is therefore due to the error introduced by the grid or histogram bin resolution. (b) The mean absolute deviation of each pixel's DEM from the group's mean DEM (mean of it's gridding group), divided by the estimated DEM inversion uncertainties for that group. Both these plots have been calculated for only DEM values above 10\% of the maximum DEM peak (maximum for each group's mean DEM), to avoid division by small numbers.}
    \label{testdem}
  \end{figure}

This type of analysis can be extended to all grid groups. For each group that contain three or more pixels, the set of directly calculated DEMs and uncertainties are recorded, and a mean DEM and uncertainty calculated (as shown as a bold line and black error bars for the example of Figure \ref{demvar}b). A test of the validity of the gridding process is the number of that group's DEMs that fit within the DEM uncertainty bounds. Figure \ref{testdem}a shows the percentage of pixels that fall outside of these bounds as a function of temperature, calculated across all 7309 groups containing three or more pixels. This value is calculated only for DEM bins above 10\% of the maximum DEM peak (for each group), otherwise the result becomes dominated by the errors at small DEM values, and division by small numbers (for example, see Figure \ref{demvar}b at higher temperatures). Only a small percentage of pixels have a part of their DEM beyond the natural uncertainty range. That is, the gridding of intensity values into bins leads to only a small increase in uncertainty. An analytical estimate of this increase is given later.

Figure \ref{testdem}b shows a similar measure - the absolute deviation of each pixel's DEM from it's group's mean DEM, divided by the group's DEM uncertainty. This is averaged across all 7309 groups containing three or more pixels, for DEM values above 10\% of the peak values (as above). This shows that the grouped DEMs tend to lie very close to their group's mean, within a range less than 30\% of the natural uncertainty on average. 

This section shows that gridding pixels into groups of similar intensities is a valid procedure for improving the efficiency of DEM inversion. Increasing the number of bins for all channels would increase the accuracy of the gridding procedure, at the expense of computational speed.

\section{DEM Gridding Uncertainty}
\label{uncertain}
The set of unique channel intensity combinations, $K_u$, is calculated. The small number of pixels not included in the range of $K_u$ are inverted directly. For each element of $K_u$, the mean intensity and measurement error is calculated for each channel (this is of course not necessary for elements of $K_u$ populated by only one measurement). The intensity and error is passed to a suitable DEM inversion method (in this case, \paperi) for processing. The resulting DEM, $D$, and uncertainty, $d$, is recorded for each $K_u$. A new error, $d^\prime$, is calculated that includes the additional error introduced due to the gridding. At a temperature bin subscripted $j$ the error is 
\begin{equation}
d^\prime_j = \sqrt{  d^2 + \sum_i^{n-1} S_{ij} \left( \Delta I_i /I_i \right) ^2 },
\end{equation}
where subscript $i$ is the index of each channel ($i=0,1,...5$), $I_i$ is the mean intensity for the current grid bin (non-logarithmic), $\Delta I_i$ is the width of the bin, and $S_{ij}$ is the relative response of each channel, $i$, at temperature bin, $j$. $S_{ij}$ is calculated from the temperature response of each channel, $R_{ij}$ (as given by the AIA Solarsoft routines) by
\begin{equation}
\label{relres}
S_{ij}=\frac{R_{ij}}{\sum_{i=0}^{n-1}R_{ij} }, 
\end{equation}
so that, at a given temperature bin, the relative responses sum to unity over all channels. See Equation 1 and related text of \paperi\ for more detail. Given a reasonable choice for the number of bins in each channel, $\Delta I_i$ is small compared to the measurement errors, and the additional error introduced by the gridding method is thus relatively small. In this paper, the choice of 16 to 26 bins depending on each channels median intensity, is a compromise between computational efficiency and the accuracy of the DEM output.

As each $K_u$ is processed, the DEM is mapped back into the original $x,y$ pixels by identifying which elements of $K$ are equal to the current value of $K_u$. All these pixels are then assigned to that DEM and the procedure is completed when all elements of $K_u$ are thus processed.

\section{Results and Comparison}
\label{results}

\begin{figure}    %%%%%%%%%%%%%%%%%% FIGURE 1 
   \centerline{\includegraphics[width=0.95\textwidth,clip=]{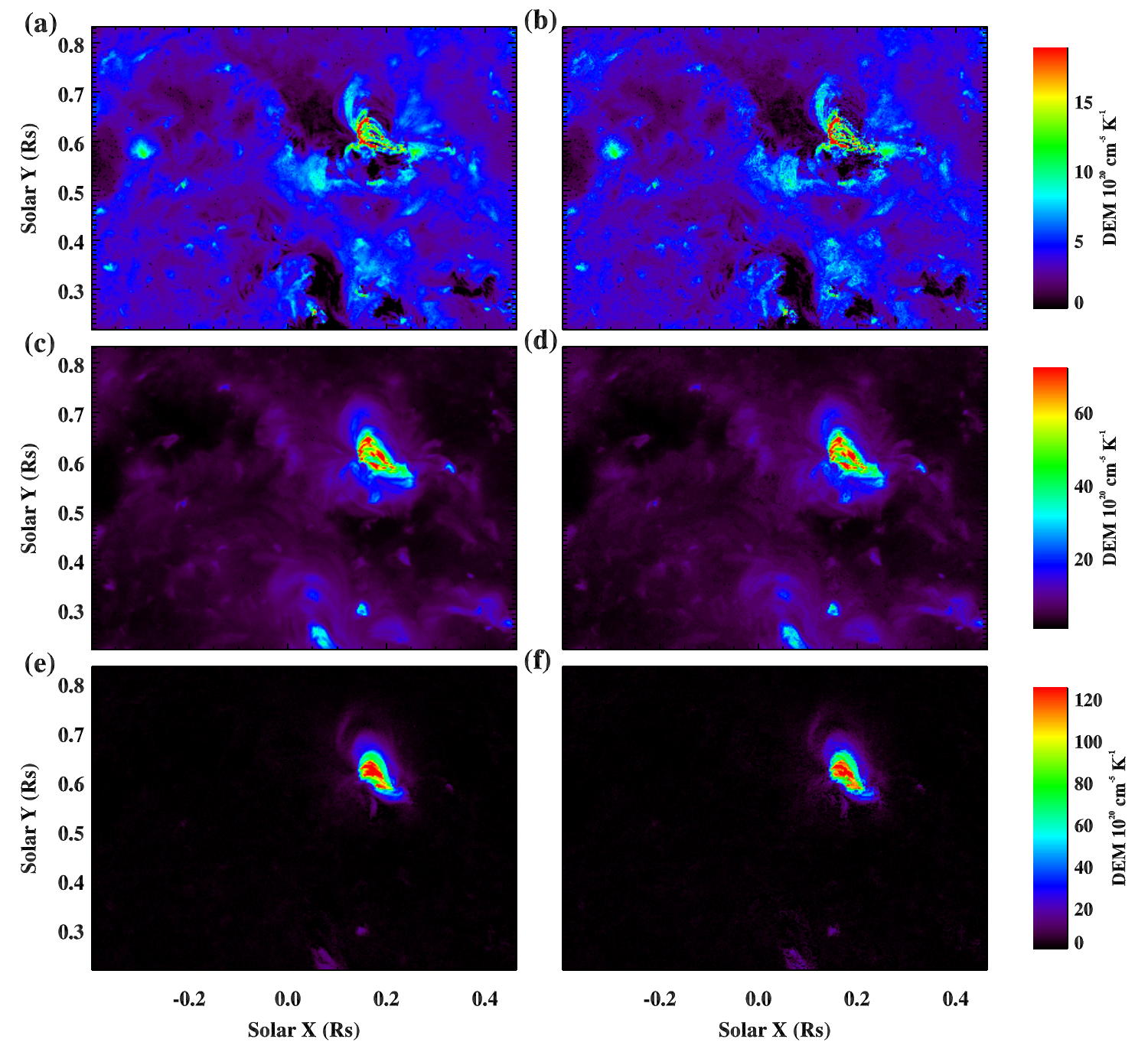}}
   \caption{Comparison of DEMs gained directly using SITES (left column) and using Grid-SITES (right column), at temperatures 1.17MK (top), 1.82MK (center) and 3.06MK (bottom).}
    \label{regiondem}
  \end{figure}

\begin{figure}    %%%%%%%%%%%%%%%%%% FIGURE 1 
   \centerline{\includegraphics[width=0.95\textwidth,clip=]{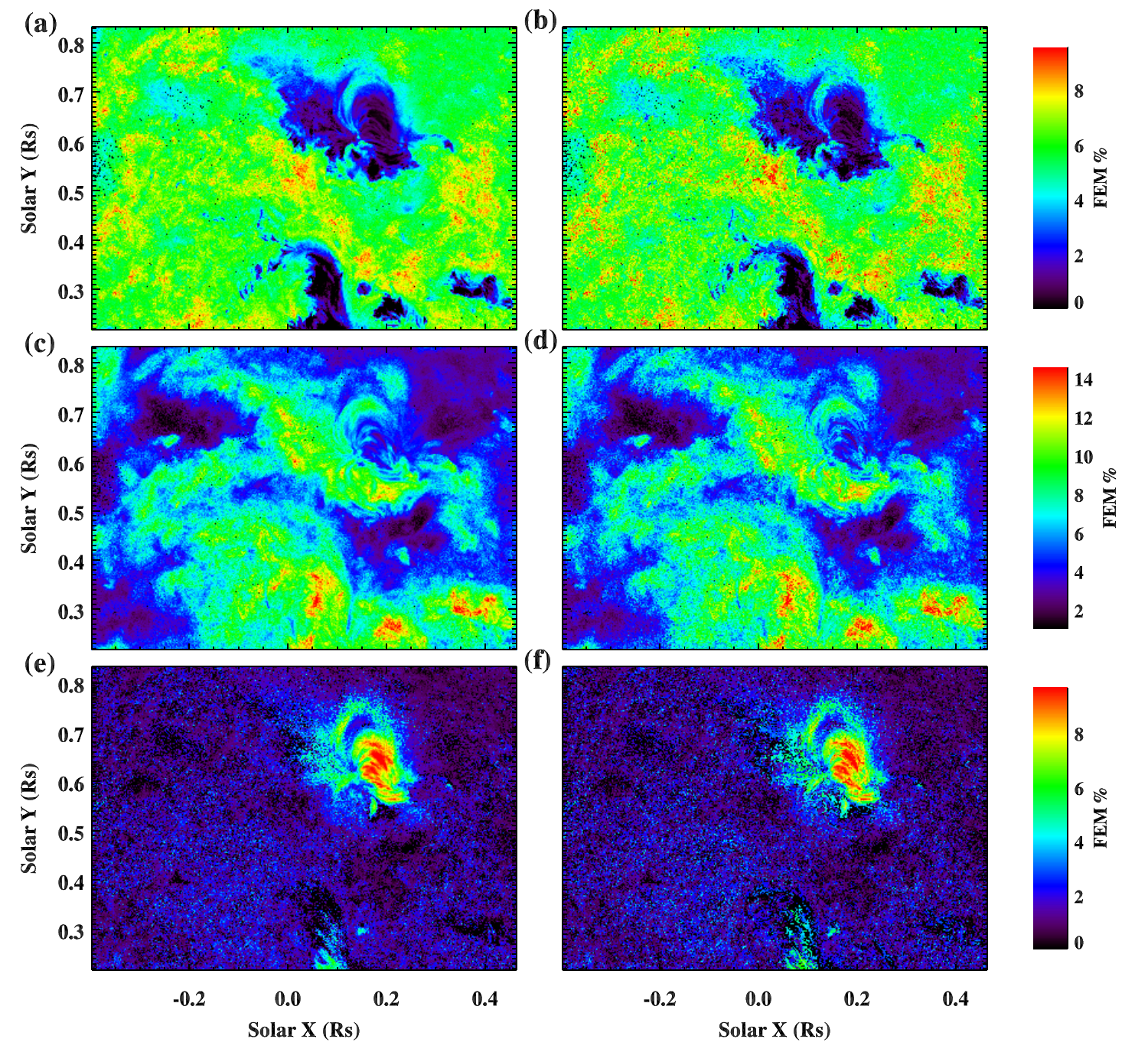}}
   \caption{Same as Figure \ref{regiondem}, but showing the fractional emission measure (FEM).}
    \label{regionfem}
  \end{figure}

This section compares DEM inversions gained directly from SITES and through the Grid-SITES method across the region of interest (boxed region in Figure \ref{aiaimage}). Figure \ref{regiondem} shows the DEM for both methods at three different temperatures. At all temperatures, the maps are very similar, appearing identical to the eye. Figure \ref{regionfem} shows the same comparison, but for the fractional emission measure (FEM). The FEM is the ratio of emission at a given temperature over the total emission integrated over all temperature at that pixel, introduced in \paperi. In this case, some minor differences can be seen. The Grid-SITES maps look slightly more pixelated and non-smooth in some isolated regions. This is due to the DEM results being quantized into discrete steps according to the gridding output.

\begin{figure}    %%%%%%%%%%%%%%%%%% FIGURE 1 
   \centerline{\includegraphics[width=0.95\textwidth,clip=]{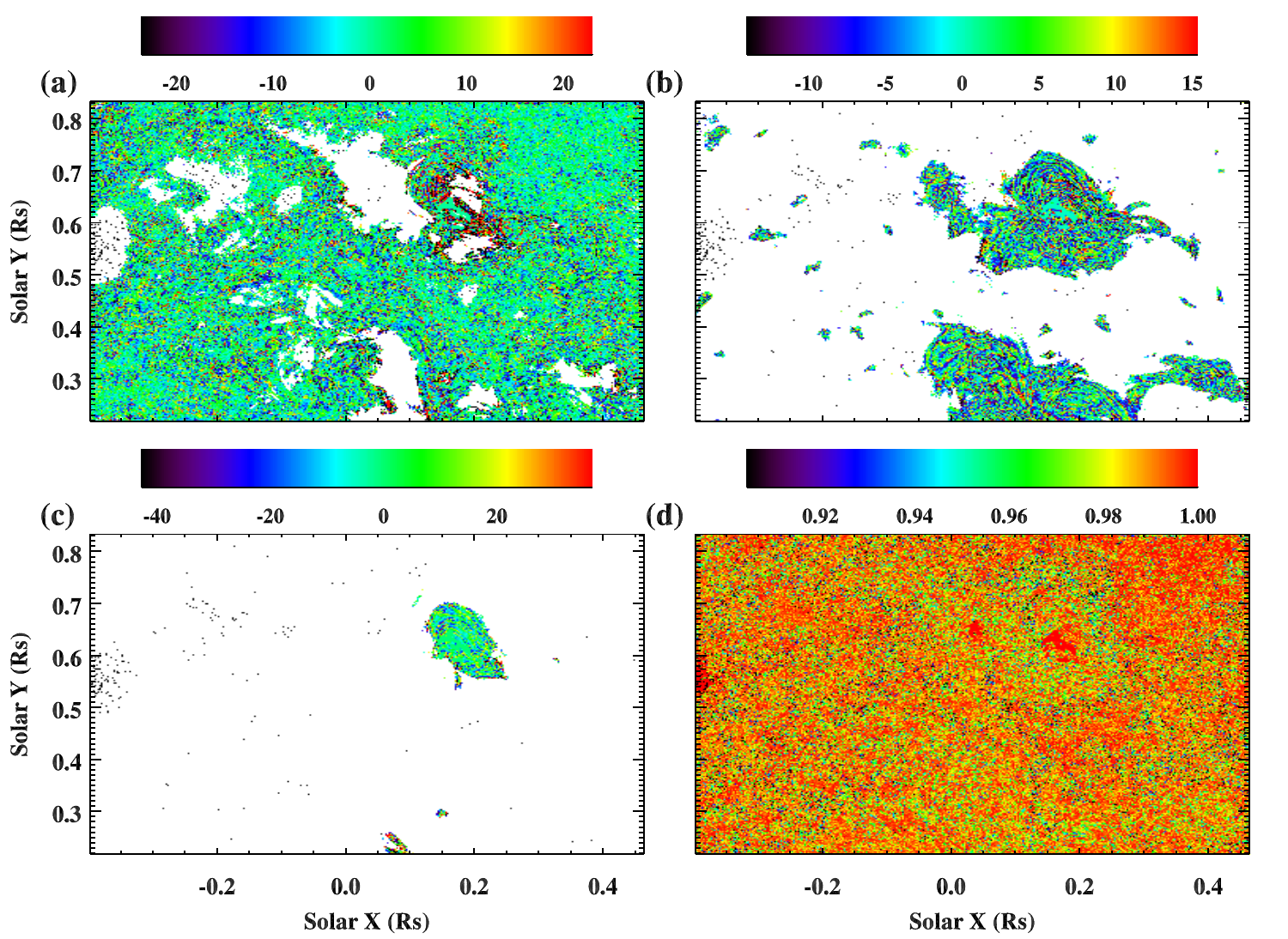}}
   \caption{(a)-(c) The percentage deviation of the Grid-SITES from the SITES DEM at temperature (a) 1.17MK, (b) 1.82MK and (c) 3.06MK. White regions are where the SITES DEM values drop below 5\% of the maximum at that temperature (see text). (d) The correlation between the SITES and Grid-SITES DEM profiles (calculated across the whole temperature range).}
    \label{regioncomp}
  \end{figure}

Figures \ref{regioncomp}a-c show the percentage deviation of the SITES and Grid-SITES DEMs at three different temperatures. The deviation, $\Delta$, is calculated as 
\begin{equation}
\Delta=\frac{D_{s}-D_g}{D_s},
 \end{equation}
 where $D_s$ is the SITES DEM, inverted directly, and $D_g$ is the Grid-SITES DEM. To avoid large relative deviations where $D_s$ is small, these values are only shown for regions where $D_s$ is larger than 5\% of the maximum $D_s$ at that temperature. The majority of pixels have small deviations, at a few percent, with only small regions or isolated pixels above $\pm$10\%. Figure \ref{regioncomp}d shows the correlation between the $D_s$ and $D_g$ profiles, calculated across the temperature range. This is a measure of how similar the DEM profiles are, regardless of their absolute values. All correlations are above 0.9, with the vast majority very close to 1.
 
The results of this section shows that Grid-SITES is a valid approach to improving the efficiency of DEM inversions. The quality of the match between directly inverted DEMs and gridded DEMs can be further improved by increasing the number of bins for each channel, at the expense of computational efficiency. 

\section{Application to Time Series}
\label{timeseries}

A highly efficient application of Grid-SITES is for processing a time series of observations. A solution grid is constructed by defining the minimum and maximum intensities, and number of bins to the first observation as described in the method above. Grid-SITES is then applied to the first observation, and the resulting DEMs recorded for each required populated bin. A new solution grid is calculated for the second observation, but using the same parameters (number of bins, minimum and maximum intensities) as the first observation. For this second observation, many of the populated grid bins will also be populated by the first observation, thus the DEMs can be immediately read from the previous grid. This avoids calculating DEMs for grid points common to both sets of observations, and only new grid points require processing, offering increased efficiency. The updated grid can then be used for subsequent observations and the process repeated, with a decreasing number of new grid points requiring DEM processing at each time step.

\begin{figure}    %%%%%%%%%%%%%%%%%% FIGURE 1 
   \centerline{\includegraphics[width=0.88\textwidth,clip=]{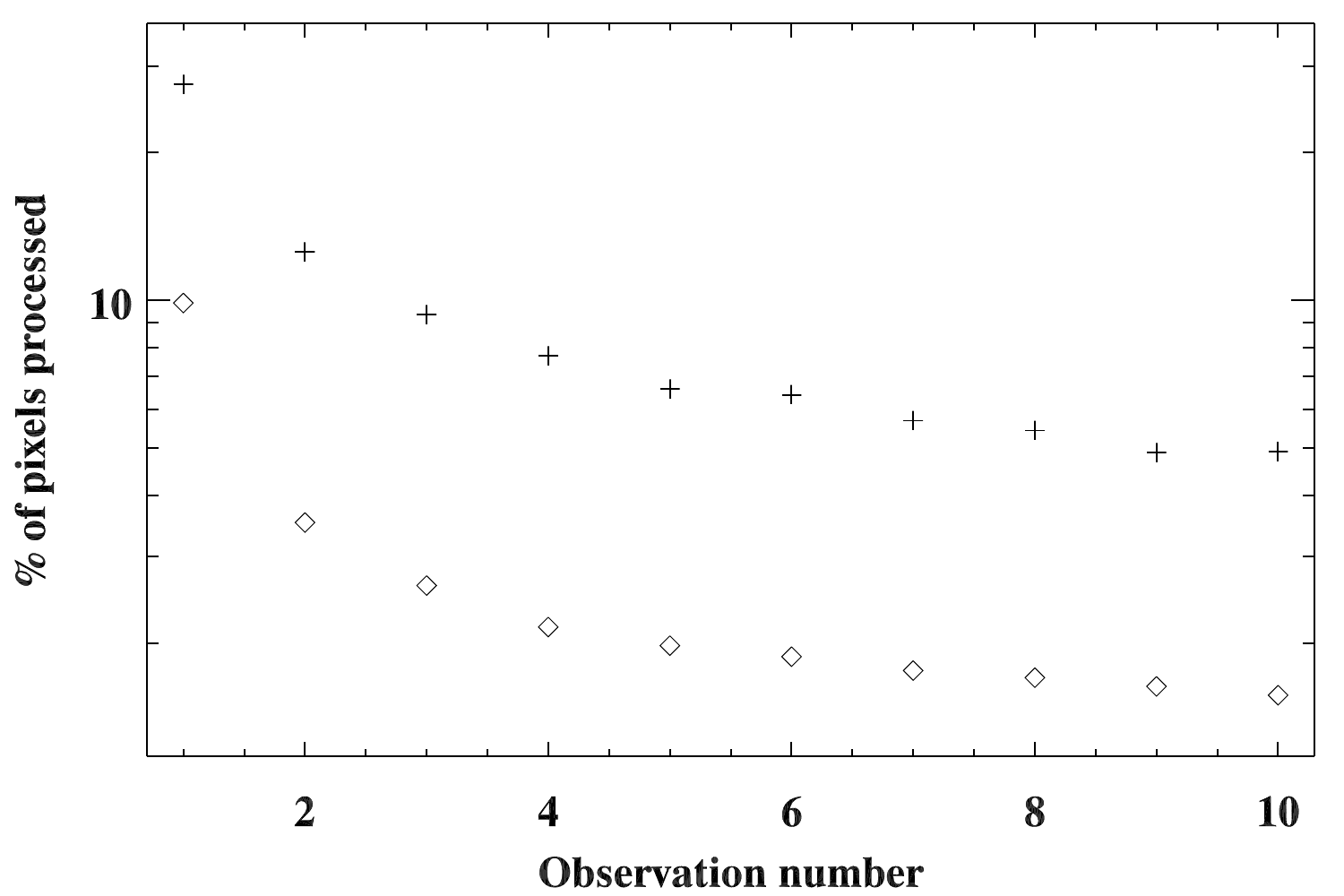}}
   \caption{The percentage of image pixels that must be processed for a time series of AIA observations, with the grid solutions maintained and updated between each set of observations (see text). The crosses are for the small example region (87918 pixels), the diamonds are for the whole solar disk (570240 pixels). A series of ten observations are processed, with the first observation at 01 January 00:00UT, with a 1 hour increment.}
    \label{timerep}
  \end{figure}

 Figure \ref{timerep} shows, for a time series of ten observations, the percentage of pixels that require processing. Results are shown for both the small example region containing 87918 pixels, and for the whole solar disk containing 570240 pixels. For the small region, the first time step requires processing of 27\% of the pixels in order to populate the initial grid. For the second time step, this drops to 12\%. By the tenth time step, only 5\% of the pixels require processing. For the whole disk, the first time step requires 10\% of pixels to be processed, dropping to 4\% by the second time step, and down to just 1.5\% by the tenth time step - thus a factor of 63 faster than processing directly.

The AIA calibration, or the temperature response curves, vary over time as the detectors degrade. This time variation is slow and linear, but has occasional discontinuous steps at certain dates, and is different across channels. Thus a Grid-SITES solution grid is only valid for a certain time period around the date for which the grid is created. During times when the calibration changes slowly and linearly, a period of one or two weeks is probably acceptable. Greater care is needed during times of discontinuous large jumps in response curves. Given this consideration, the DEM solution grid can be saved, and used for processing of other datasets as long as they are collected within an acceptable time period. Care must also be taken with a time series of observations containing rapid changes in intensities, for example during a flare or a coronal dimming. In this case, the intensity ranges for each channel that form the initial grid parameters should be set at ranges appropriate for the whole series. The AIA data file headers contain useful DATAMIN and DATAMAX fields.

\section{Application to Full-disk Images}
\label{fulldisk}

\begin{figure}    %%%%%%%%%%%%%%%%%% FIGURE 1 
   \centerline{\includegraphics[width=0.6\textwidth,clip=]{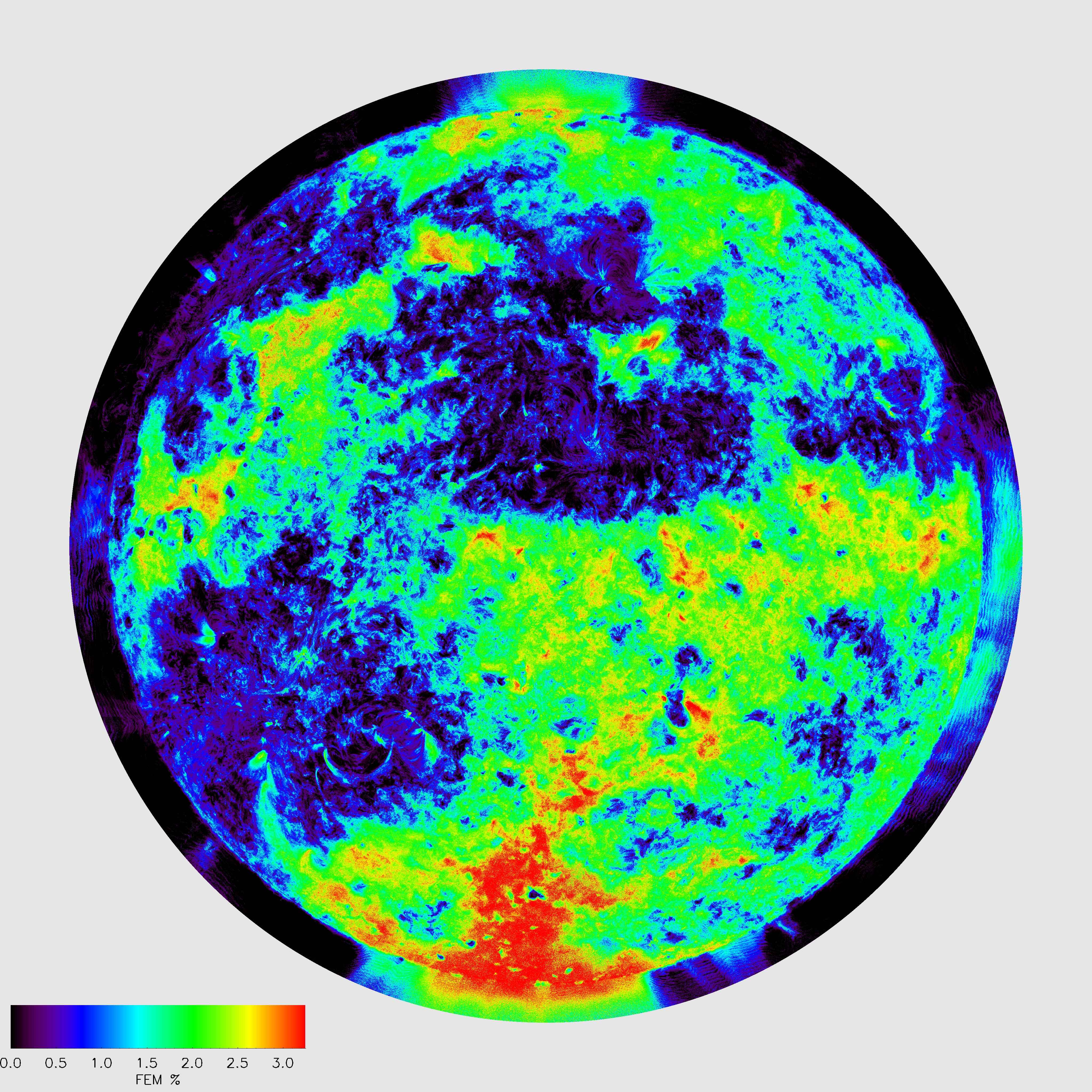}}
      \centerline{\includegraphics[width=0.6\textwidth,clip=]{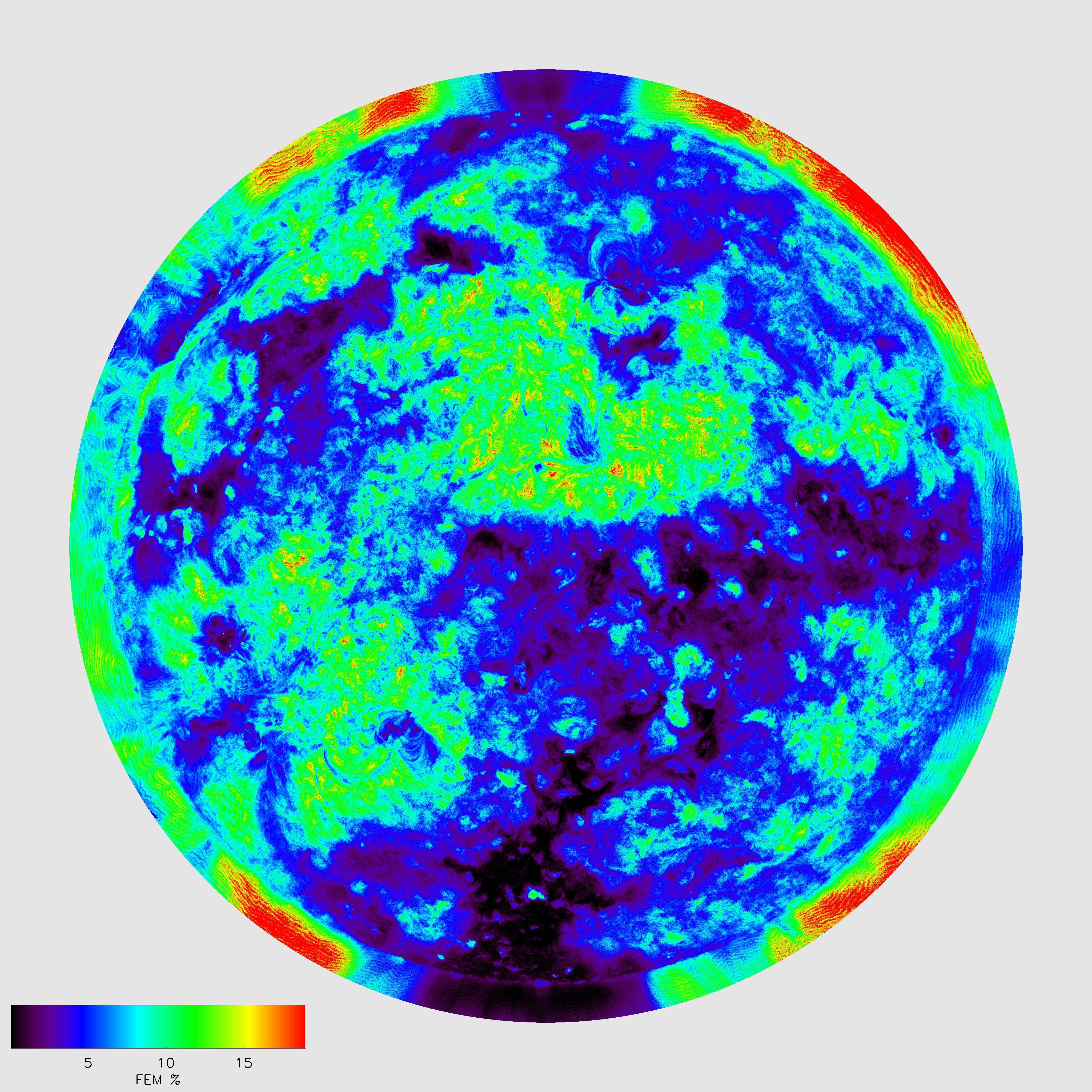}}
         \centerline{\includegraphics[width=0.6\textwidth,clip=]{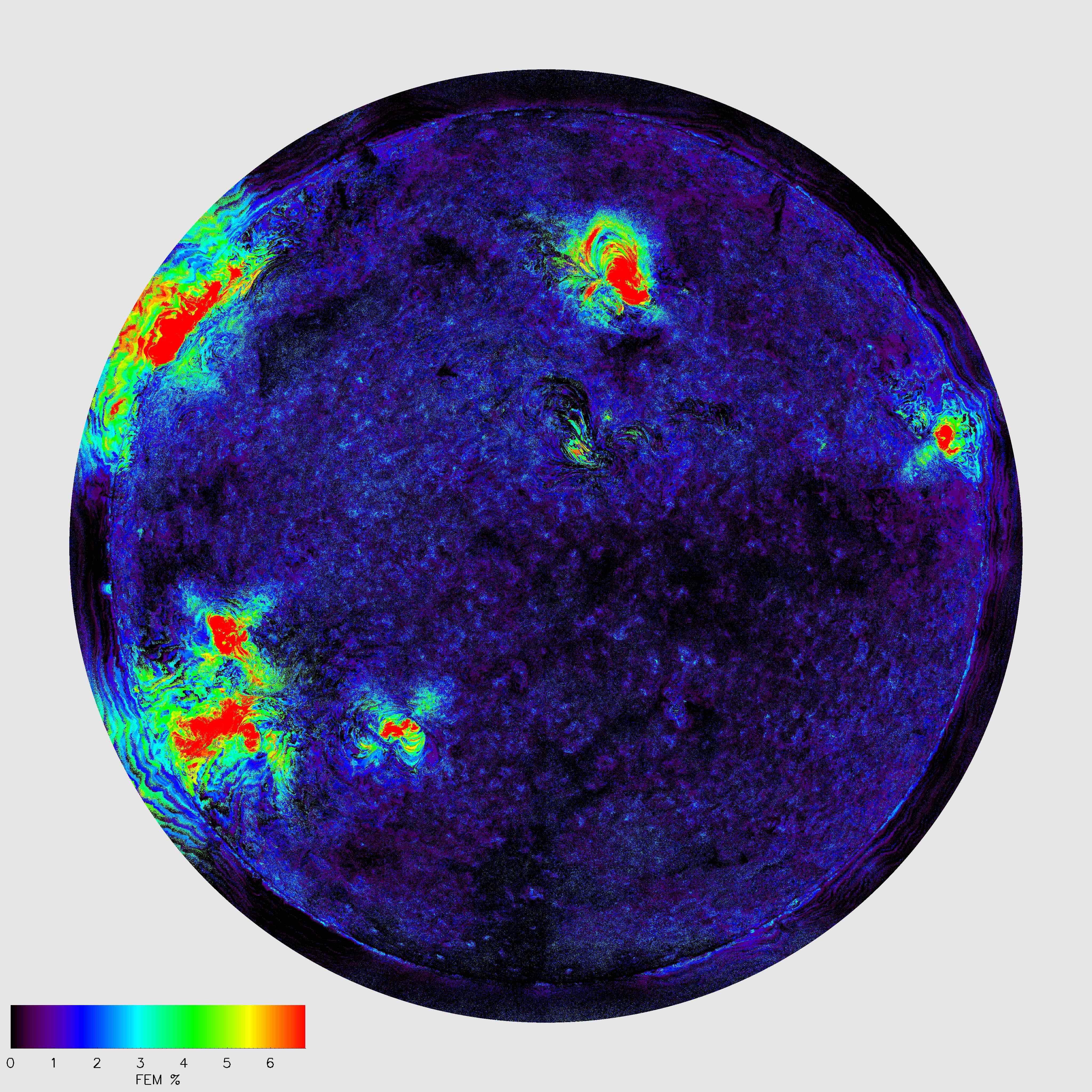}}
   \caption{Fractional emission measure (FEM) at a temperature of 0.5MK (top), 1.6MK (middle) and 3.1MK (bottom) for a full-resolution observation made at 01 January 00:00UT, and inverted using Grid-SITES. The colour bar shows the percentage FEM for each pixel at each temperature.}
    \label{fem}
  \end{figure}

Figure \ref{fem} shows the FEM resulting from applying Grid-SITES to a full-resolution whole disk observation of 01 January 00:00UT (the same observation is visualised in Figure \ref{aiaimage}). Three temperatures are shown: 
\begin{itemize}
\item 0.5MK: This is at the lower limit of valid temperatures for SITES inversion (see \paperi). The FEM at this temperature is dominated by coronal holes and the quietest Sun far from active regions.
\item 1.6MK: At this temperature, coronal holes, the quietest Sun, and active regions have low FEM. The FEM is dominated by broad regions surrounding active regions. The high FEM values in these regions ($\approx$15\%) indicate DEMs that peak strongly near 1.6MK.
\item 3.1MK: Only the active regions have high FEM at this temperature.
\end{itemize}

 The full-disk observation of Figure \ref{fem} has just over $10^7$ pixels in the processed field of view. Grid-SITES finds $3.33 \times 10^5$ groupings of pixels which gives a 30-fold increase in computational efficiency. This observation was processed in around ten minutes on a 3.7GHz Intel Core i5 desktop PC with 32Gb memory, compared with 6 hours without Grid-SITES. If the calculated grid was reused for other observations made close in time, full-disk images may be processed in less than a minute.
 
 In high-resolution Grid-SITES DEM (or FEM) maps, the quantisation of the input data (and output DEMs) can be seen as discontinuous steps in off-limb regions. This is seen particularly clearly above limb active regions in the bottom panel of Figure \ref{fem}. One possible improvement would be an interpolation scheme to avoid these discrete steps. The interpolation is not straightforward since it must be done in the multi-dimensional space of the input data grid, and the populated grid bin may have neighbouring bins that are empty, thus contain missing information for interpolation. In principle, if a working interpolation scheme is achieved, the Grid-SITES DEMs should more closely match directly-inverted DEMs.

\section{Summary}
\label{summary}

Grouping similar groups of pixels according to their logarithmic intensity is a valid approach to improve the efficiency of DEM inversions for large sets of data. For small images (\emph{e.g.} $\approx10^5$ pixels) the efficiency increase is $\approx5$-fold, for larger images or time series of images (\emph{e.g.} $\approx10^7$ pixels) the efficiency is closer to $100-$fold. The gridding leads to output DEMs that agree well with directly-inverted DEMs, with only a small relative increase in uncertainty. 

The accuracy of Grid-SITES is dictated by the choice of grid resolution. Finer grids (higher number of bins) lead to higher accuracy and less efficiency. In this paper, 16-26 bins are allocated to each channel according to that channel's typical intensities, giving a sensible compromise between accuracy and efficiency. In the limit of very fine grids, the result and efficiency becomes equal to inverting the data directly per pixel. Thus Grid-SITES offers a flexible working environment according to the purpose of the inversion. For example, a user can choose a coarse binning for a very fast inversion to give initial results, and a fine grid for more detailed further analysis. Greater efficiency gains are offered by larger datasets, and a grid can be saved for use on other data as long as the user is aware of certain periods when the AIA response calibration changes rapidly. 

In this work, Grid-SITES uses SITES as the core inversion method, although the same gridding approach can be used with any DEM inversion method. This can facilitate the application of more than one inversion method to the same dataset, giving additional confidence on the inversion results. This work has used AIA images exclusively, although Grid-SITES can obviously be used for inverting any suitable dataset. Furthermore, the gridding scheme can be used for analysis other than DEMs, where a time-consuming procedure is to be applied to a large multi-dimensional dataset.

The incentive for developing Grid-SITES is to analyse large datasets and enabling, for example, large-scale statistical studies of active regions, or of coronal changes over long time-scales using AIA/SDO. Given a decent desktop computer, Grid-SITES enables rapid DEM inversions of a time series of full-resolution, full-disk AIA observations - in minutes rather than hours. The IDL routines for Grid-SITES, the DEM inversion SITES method, plus the FEM visualisation method will be made available via the Solarsoft software package. 

\begin{acks}
We thank an anonymous referee for comments that improved this article. James Pickering is supported by an STFC studentship. Part of Huw Morgan's work on this project is supported by an STFC consolidated grant to Aberystwyth University. CHIANTI is a collaborative project involving George Mason University, the University of Michigan (USA), University of Cambridge (UK) and NASA Goddard Space Flight Center (USA). 

Conflict of Interest: The authors declare that they have no conflict of interest.

\end{acks}

\bibliographystyle{spr-mp-sola}
% name your Bibtex file containing your references (.bib)
\bibliography{./biblio.bib}

\end{article} 

\end{document}